\documentclass{article}

\usepackage{ragged2e}
\usepackage{graphicx}
\usepackage{caption}
\usepackage[margin=0.8in]{geometry}
\usepackage{fancyhdr}
\usepackage{gensymb}
\usepackage{hyperref}
\RaggedRight

\graphicspath{{./figs/}}  

\makeatletter
\renewcommand{\maketitle}{
\huge
\@title
\medskip
\global\@topnum\z@
}
\makeatother 
   
 \renewenvironment{author} {
 \large
 \emph
  }
j 

\newcommand{\keywords}[1]{
\medskip
Keywords: \textit{#1}
}

\newcommand{\dedication}[1]{
\medskip
\textit{#1}
}

\newenvironment{affiliations}{
\medskip
\large
}

\renewenvironment{abstract}{
\small
\medskip\medskip
}

\newcommand{\threesubsection}[1]{
\medskip
\textit{#1}: 
}

\def\Re{\mathop{\mathrm{Re}}}
\def\Im{\mathop{\mathrm{Im}}}

\usepackage{color}
\usepackage{amsmath,amssymb}
\begin{document}
\pagestyle{fancy}

\title{
Dynamic quarter-wave metasurface for efficient helicity inversion of polarization
beyond the single-layer conversion limit
}

\maketitle


\author{Mitsuki Kobachi}
\author{Fumiaki Miyamaru}
\author{Toshihiro Nakanishi}
\author{Kunio Okimura}
\author{Atsushi Sanada}

\author{Yosuke Nakata*}


\dedication{}

\begin{affiliations}
M. Kobachi, Prof.~A. Sanada\\
Graduate School of Engineering Science, Osaka University, Osaka 560-8531, Japan

Prof.~F. Miyamaru\\
Department of Physics, Faculty of Science, Shinshu University, Nagano 390-8621, Japan

Dr.~T. Nakanishi\\
Department of Electronic Science and Engineering, Kyoto University, Kyoto 615-8510, Japan

Prof.~K. Okimura\\
School of Engineering, Tokai University, Kanagawa 259-1292, Japan

Dr. Y. Nakata\\
Graduate School of Engineering Science, Osaka University, Osaka 560-8531, Japan\\
Center for Quantum Information and Quantum Biology, Osaka University, Osaka 560-8531, Japan\\
Email Address: nakata@ee.es.osaka-u.ac.jp

\end{affiliations}



\begin{abstract} 
Terahertz chiral sensing and polarization-multiplexing communication demand 
subwavelength devices that dynamically invert polarization helicity.
Metasurfaces can enhance anisotropy and fine tunability at subwavelength scales for this purpose. Although metasurfaces enabling deep modulation between orthogonal polarizations have been designed, they suffer from low conversion efficiencies. In this study, it is shown that the efficiency of conversion
from linear to circular polarization by conventional single-layer transmissive metasurfaces cannot exceed a fundamental limit.
A dynamic reflective metasurface free from this limitation is then proposed.
The device includes multilayer structures working as a terahertz quarter-wave plate with switchable slow and fast axes.
A phase transition of vanadium dioxide induces the necessary structural transformation of the metallic patterns.
A practical fabrication method, based on the room-temperature bonding technique of sapphires, is presented.
Dynamic helicity inversion is demonstrated at $0.90\,\mathrm{THz}$, with a conversion efficiency of over 80\,\%
that is beyond the fundamental limit of single-layer transmissive metasurfaces ($65\,\%$)
and more than four times greater than that of previously reported devices.
\end{abstract}


\section{Introduction}

Polarization is one of the most fundamental properties of electromagnetic waves. 
It plays a crucial role in applications such as liquid-crystal displays, polarization-sensitive spectroscopy, and polarization-multiplexing communication.
As polarization control demands an anisotropic response, 
birefringent materials have been widely employed in static polarization devices, such as polarizers and wave plates.
However, the anisotropy of natural materials is typically small and limited to fixed values.
Polarization devices based on birefringent materials involve a thickness much larger than the working wavelength.
Therefore, they become bulky, especially in the lower frequency (e.g., microwave and terahertz) range.

Artificial structural surfaces called metasurfaces can be leveraged to enhance anisotropy and make devices compact \cite{Chen2016,He2018}.
They can exhibit a considerable anisotropic response even at subwavelength thicknesses, when metal or dielectric resonators are introduced.
As this response originates in anisotropic structural resonances,
it can be finely tuned by appropriate design of the structures on the metasurface.
Thus, metasurfaces can provide both larger anisotropy and wider tunability for polarization control than can natural materials, the characteristics of which are determined by the atomic constituents.
A variety of static-polarization-control metasurfaces have been developed in the microwave-to-optical frequency range, including static metasurface polarizers \cite{Takano2011,Suzuki2016} and wave plates \cite{Grady2013, Ding2015, Nakata2017, Han2018, Liu2019a, You2020}. 
However, for sensitive measurement and communication technologies, dynamic modulation of polarization is an essential requirement.
Therefore, considerable effort has been devoted to developing metasurface modulators \cite{Herrmann2020,Konishi2020}.
In the microwave region, varactor diodes have been used to realize tunable phase shifts \cite{Zhu2013} and 
dynamic polarization rotations of arbitrarily polarized incident waves \cite{Wu2019}.
A power-dependent metasurface switchable between a quarter-wave plate and a 
spatial wave manipulator has also been proposed \cite{Kiani2020}.
Even in the terahertz region, a high-electron-mobility transistor (HEMT) has been used to produce a dynamic phase shift \cite{Zhang2018};
graphene has also been applied to a gate-tunable THz polarizer \cite{Miao2015}.
Photoexcitation of semiconductors could enable ultrafast terahertz polarization modulation using gratings \cite{Okada2010, Yang2017}
and chiral structures \cite{Kanda2009, Zhang2012, Kanda2014}.

For continuous polarization tuning, micro-electro-mechanical systems (MEMS) have been applied to polarization modulators with chiral structures \cite{Kan2013, Kan2015};
polarization converters with tunable cross-conversion rates \cite{Zhang2017};
tunable wave plates that can modulate a transmitted wave from pure circular to linear polarization \cite{Zhao2018a};
and kirigami polarization modulators \cite{Choi2019}.
Phase-changeable materials provide another approach to polarization control, advantageous for deep modulation.
For example, vanadium dioxide has been employed in metasurfaces \cite{Liu2018a},
because it can induce a high-contrast resistance change over the factor of $10^3$ in phase change \cite{Okimura2005}
that has been used to alter critically the operating frequencies of THz quarter-wave plates \cite{Wang2015b, Wang2016}.
In contrast to simple operating-frequency modulation, polarization modulation at a specific frequency
requires more complex designs of structures working for both states before and after the phase transition.
Dynamic modulation between linear and circular polarizations has been demonstrated by using I-shape structures \cite{Nouman2018a}.
To realize deeper polarization modulation, checkerboard structures have been utilized in
switchable linear polarizers with a high extinction ratio \cite{Nakata2016} and helicity-switchable quarter-wave plates
that achieve orthogonal-polarization modulation \cite{Nakata2019c, Nakanishi2020}.
However, these vanadium-dioxide polarization devices, based on single-layer designs, suffer from a low conversion efficiency for incident waves,
attributable to
reflection loss from both substrate and metasurface.
Substrate Fresnel reflection could be reduced by using 
an extremely thin substrate (although implementing this is challenging), but reflection loss from the metasurface is unavoidable even in principle \cite{Ding2015}.
These two factors prevent the simultaneous achievement of orthogonal-polarization modulation and high efficiency, especially in terahertz or higher frequency regions.
Although multilayer vanadium-dioxide metasurfaces have been seen as promising for use in efficient multifunctional devices, such as 
reconfigurable absorbers \cite{Ding2018, Song2018,  Li2019, Wang2019, Song2020a, He2020, Song2020, Zhang2020, Li2020, Zhang2021a, Ren2021, Liang2021, Liu2021a}, reprogrammable  wavefront-engineering metasurfaces \cite{Shabanpour2020, Ren2021a},
and devices switchable between quarter- and half-wave plates \cite{Luo2020},
these proposals have
not been realized experimentally, because of the stringent fabrication challenges.
Although a few studies have reported experimental realization of 
multilayer vanadium-dioxide structures for infrared polarization modulation \cite{Jia2018} and terahertz asymmetric transmission \cite{Liu2019},
they did not focus on orthogonal-polarization modulation.

In this study, we experimentally demonstrate dynamic helicity inversion
with high conversion efficiency in the terahertz region
necessary for chiral sensing and communication technologies.
As mentioned above, transmissive metasurfaces are known empirically to suffer from low conversion efficiencies.
We begin by deriving a theoretical limit for the conversion efficiency of single-layer transmissive metasurfaces 
between linear and circular polarization.
We then propose a dynamic reflective metasurface with transformable metallic structures capable of overcoming this limitation.
As shown in \textbf{Figure~\ref{fig:principle}}a and b, the device efficiently converts a linearly polarized wave into a reflected circularly polarized one with switchable helicity. 
The device works as a dynamic quarter-wave plate with exchangeable slow and fast axes.
The structure is designed to achieve orthogonal-polarization modulation from the incident linearly polarized radiation with high conversion efficiency.
To resolve the fabrication challenges of multilayer structures,
we developed a fabrication method based on sapphire room-temperature bonding techniques \cite{Shimatsu2010} that is compatible with the vanadium-dioxide process.
Finally, we realize the proposed device physically and demonstrate its highly efficient helicity-switching function in the terahertz frequency range. 
Note that we use the convention of a harmonic time dependence $\exp(j \omega t)$ with imaginary unit $j$ in this study.

\section{Results and Discussion}

\subsection{Helicity conversion efficiency limit of single-layer metasurfaces}
First, we define a figure of merit for helicity conversion using quarter-wave plates.
Consider a plane wave propagating in the $z$ direction and normally incident on a quarter-wave plate with fast and slow axes along the $x$ and $y$ axes (or vice versa).
As the quarter-wave plate can convert diagonal ($y=x$) linear polarization ($D$ polarization) into a circular one,
we assume that the incident wave has polarization $D$.
In a previous paper \cite{Nakata2019c}, the $z$ component of the Stokes parameter $S_3$ was divided by
the transmitted power flux $S_\mathrm{out}$ to define a circular-polarization ratio 
\begin{equation}
 \frac{S_3}{S_\mathrm{out}} = \frac{2\Im (\tilde{t}_{x}^*\tilde{t}_y)}{|\tilde{t}_x|^2 + |\tilde{t}_y|^2},  \label{eq:1}
\end{equation}
where $\tilde{t}_i$ ($i=x,y$) is the complex amplitude transmittance of the $i$ polarization.
Equation~(\ref{eq:1}), which characterizes the purity of the output helicity state converted from $D$-polarization, can be used to evaluate the polarization-state modulation rate.
Here, $S_3/S_\mathrm{out}= \pm 1$ corresponds respectively to a pure right- or left-circularly polarized wave when viewed from the receiver.
We next introduce the helicity-conversion efficiency
\begin{equation}
\frac{S_3}{S_\mathrm{in}} =  \Im (\tilde{t}_{x}^*\tilde{t}_y),  \label{eq:2}
\end{equation}
where $S_\mathrm{in}$ is the incident-beam power flux.
$S_3/S_\mathrm{in}= \pm 1$ represents a perfect dissipation-free conversion from linear polarization to 
right- or left-circular polarization, respectively.
We can use $S_3/S_\mathrm{in}$ as a figure of merit that includes both polarization and efficiency information.

Next, we derive the fundamental maximum limit of $|S_3/S_\mathrm{in}|$ for single-layer transmissive metasurfaces.
In contrast to the avoidable substrate Fresnel loss, metasurface reflection loss is inevitable.
Consider a plane wave normally incident on a transmissive electric metasurface at $z=0$ in a vacuum.
The metasurface has an infinitesimally small thickness in the long-wavelength approximation, 
where the Fresnel reflection of the substrate is negligibly small.
For simplicity, we assume that the incident frequency is lower than the diffraction frequency, and the metasurface has mirror symmetry with respect to either $x=0$ or $y=0$.
Then, there is no diffraction into higher-order modes
and no cross conversion between $x$ and $y$ polarizations:
the metasurface behaves as a sheet characterized by different complex-amplitude transmittances $\tilde{t}_x$ and $\tilde{t}_y$.
From the boundary conditions of electric-field continuity 
and energy conservation, the complex transmittance must satisfy the constraint
\begin{equation}
 |\tilde{t}_i |^2+|1-\tilde{t}_i |^2=1 \quad (i=x,y),
\end{equation}
i.e., the transmittance is constrained at the circumference of a circle 
with 1/2 center and 1/2 radius in the complex plane.
High transmission $|\tilde{t}_i|=1$ only occurs when $\arg (\tilde{t}_i)=0$;
a large transmission phase leads to transmission deterioration. 
From this restriction, the helicity-conversion efficiency $|S_3/S_\mathrm{in}|$ 
has its maximum of $3\sqrt{3}/8 \approx 0.65$ for $\tilde{t}_x=(1+\exp(\pm j \pi /3))/2$ and $\tilde{t}_y=(1+\exp(\mp j \pi /3))/2$.
Thus, single-layer transmissive metasurfaces have a fundamental efficiency limitation
even if we eliminate substrate reflection.

\subsection{Design principles of multi-layered device}
To overcome the limitation described above, 
we propose a reflective metasurface on a dielectric substrate with a metal ground.
In contrast to transmissive devices, a reflective device does not have unwanted scattering channels 
that cause dissipation. 
Thus, a reflective metasurface allows 
perfect polarization conversion in an ideal case without the fundamental limitation
of single-layer transmissive devices and with much higher efficiency. 
Note that single-layer reflective converters with prisms have already enabled non-orthogonal modulation between
linear and circular polarization, but they are bulky and exhibit prism reflection \cite{Liu2018}. 
Thus, we should consider a multilayer reflective device.

\begin{figure}
  \includegraphics{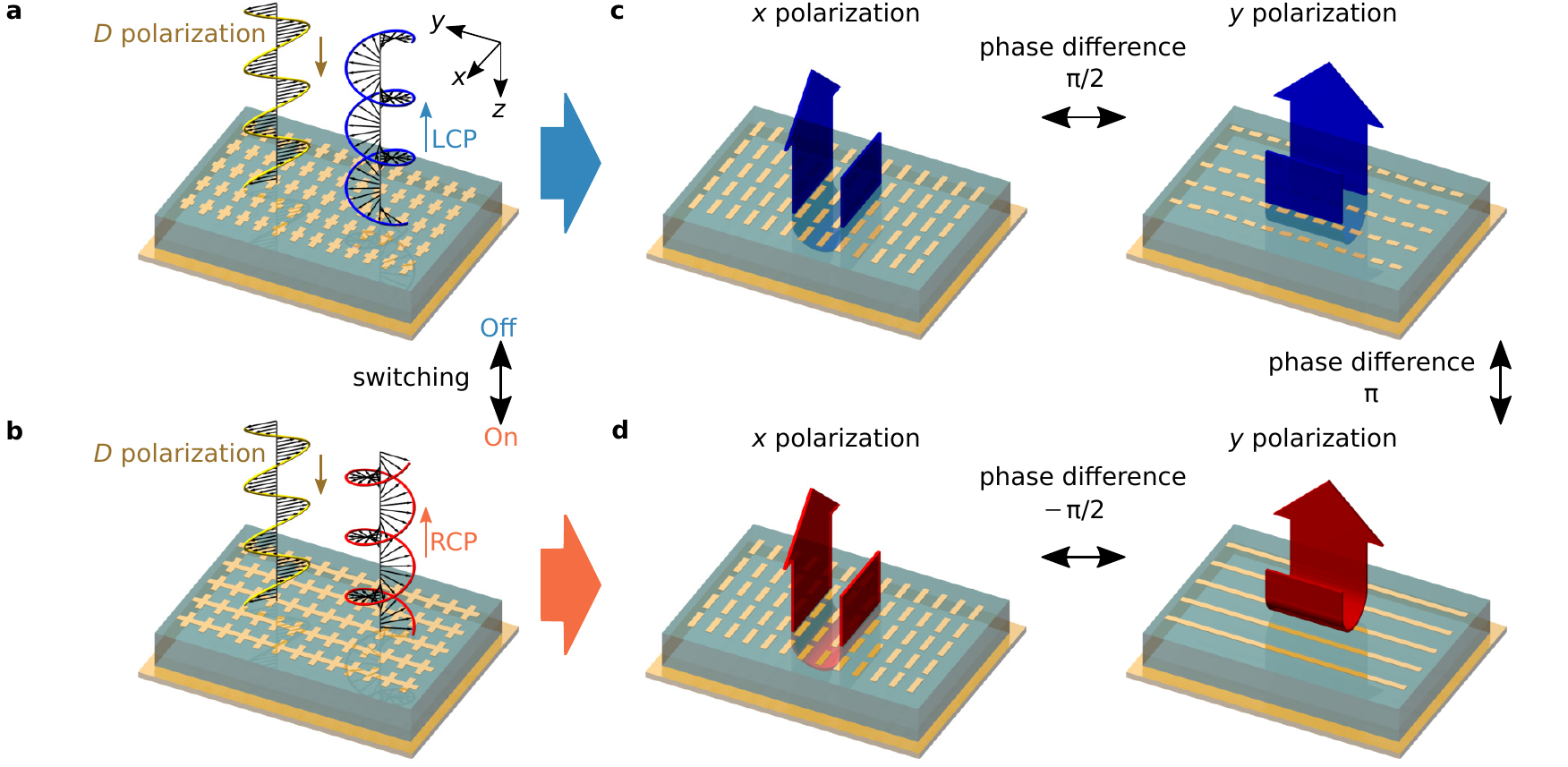}
  \caption{a), b) Schematic of a dynamic reflective quarter-wave metasurface for circular-polarization switching.
The metallic structures are transformed when the device is switched a) off or b) on.
The incident linear ($D$) polarization is converted to 
left-circular polarization (LCP) or right-circular polarization (RCP),
depending on the device state.
c), d) Design principles of a dynamic reflective quarter-wave metasurface.
The separately-designed responses to $x$ and $y$ polarizations for c) off and d) on states are combined to realize the helicity-switching function. 
 }
  \label{fig:principle}
\end{figure}

To implement the necessary structural transformation, we utilize vanadium dioxide, which
exhibits a metal--insulator transition at a critical temperature of $T_c\approx 340\,\mathrm{K}$ \cite{Nag2008}.
The transition involves change of vanadium-dioxide crystalline structures,
but the resistivity variation is stable under multiple cycles of transitions \cite{Ko2008}.
When the device is off, it is in its low-temperature (insulating) state; it is in its high-temperature (metallic) state when on.
To convert $D$ polarization into circular polarization, a phase shift $\pm \pi/2$ between the $x$ and $y$ polarizations is necessary.
When we assume mirror symmetry, 
the device independently responds to $x$ and $y$ polarizations, 
which may therefore be treated separately.
First, we consider an off-state device, as shown in Figure~\ref{fig:principle}c.
Because the metallic patches are shorter than the target wavelength,
most of the incident $x$- and $y$-polarized waves are transmitted through the dielectric surface 
and reflected back at the ground plane. 
To induce a phase shift $\pi/2$ between polarizations,
the lengths of the metallic structures are tuned. 
Second, we consider the on-state device [Figure~\ref{fig:principle}d].  
The temperature change induces a structural deformation along the $y$ axis from metal patches to the wire grid.
The $y$-polarized wave of wavelength $\lambda$ is now reflected at the surface due to the
wire-grid response \cite{Takano2011}.
The thickness of the dielectric substrate is set to approximately $\lambda/4$; this
induces a round-trip phase shift of $\pi$ radians between the off and on states.
Combining the structures in the $x$ and $y$ directions,
we can realize helicity switching, as shown in Figure~\ref{fig:principle}a and b.
When a normally incident wave with $D$ ($y=x$) polarization enters the off-state device, the
$\pi/2$ phase difference between the $x$ and $y$ polarizations 
converts the $D$ polarization into left-circular polarization.
When the device is turned on,
the additional $\pi$-radian phase shift is added only to the
$y$ polarization,
leading to helicity inversion of the reflected wave. 

\subsection{Simulation}
\begin{figure}
  \includegraphics{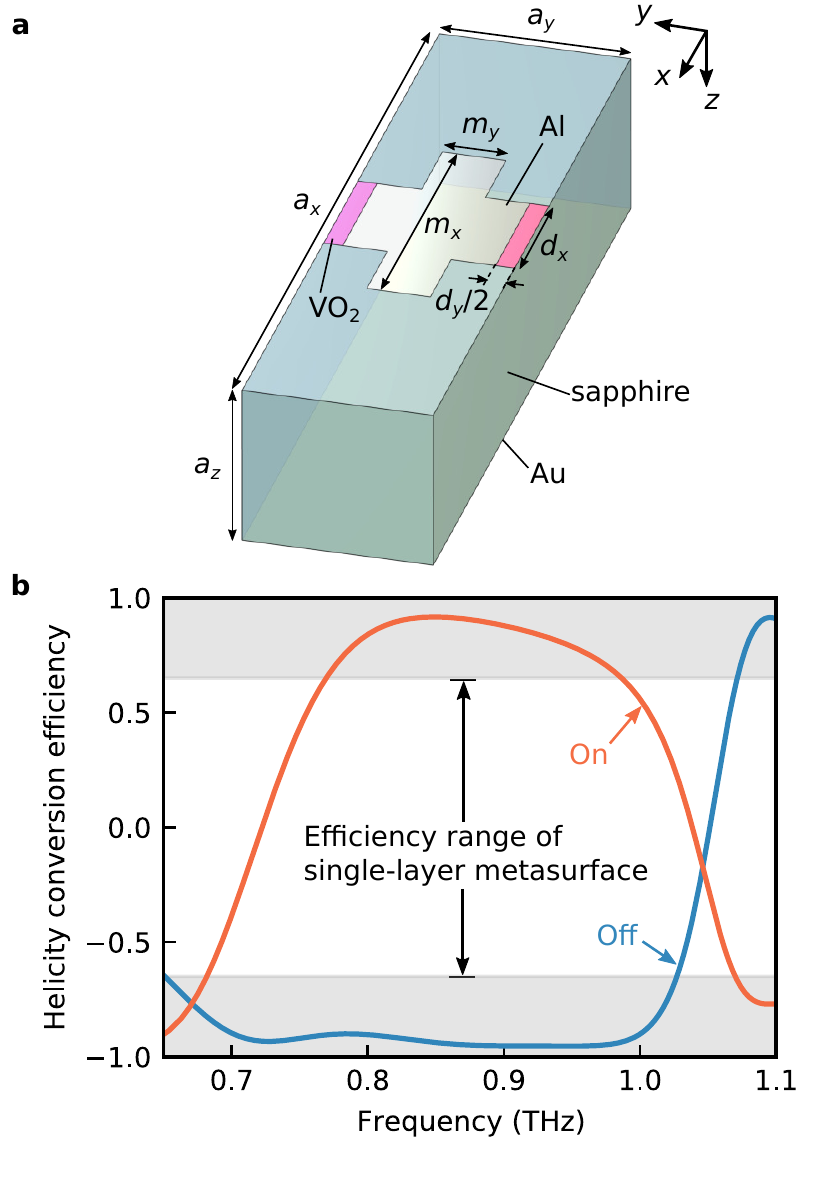}
  \caption{ a) Unit-cell design of the device, showing definitions of dimensional parameters. b) Calculated helicity-conversion efficiency, $S_3/S_{\mathrm{in}}$ of proposed device. In the gray area, the efficiency exceeds the limiting efficiency of single-layer transmissive devices. }
  \label{fig:simulation}
\end{figure}

To design the device structures, we use an electromagnetic-field simulator \textsc{COMSOL Multiphysics}. 
\textbf{Figure~\ref{fig:simulation}}a shows the device unit cell.
Exploiting of the mirror symmetries, we reduced computational costs by performing the calculation only for a quarter part of the unit cell with appropriate boundary conditions.
The device is placed in vacuum and has a sapphire dielectric
substrate;
its refractive indices are 
$n_{x,y}=3.1 - 0.0021j$ and $ n_{z}=3.4 - 0.0023j$ \cite{Grischkowsky1990}.
The sapphire orientation is chosen so that the c plane lies on the surface, 
where the vanadium dioxide grows finely because of lattice
matching \cite{Nag2008}.
On the bottom of the sapphire substrate, 
we place a gold ground plane, which is represented in the simulation by an impedance boundary
with an effectively infinite thickness. 
Aluminum and vanadium-dioxide patterns are formed on the top of the
c-plane substrate.
The aluminum region is treated as a transition boundary
with a thickness of $400\,\mathrm{nm}$.
The conductivities of gold and aluminum are
$31\,\mathrm{S/\mu m}$ and
$22 \,\mathrm{S/\mu m}$, respectively \cite{Laman2008}.
The vanadium dioxide is represented by a sheet resistance of 
$500\,\mathrm{k\Omega}/\square$ and $10\,\mathrm{\Omega}/\square$ for the off
and on states, respectively.
The $x$- and $y$-polarized waves impinge on the surface normally, and 
complex reflection amplitudes are calculated.
We set the substrate thickness 
$a_z$ to $26\,\mathrm{\mu m}$
to induce a $\pi$ phase difference between the off 
and on states for a $y$-polarized wave at $0.93\,\mathrm{THz}$.
Next, we consider the structures, described by
the dimensional parameters defined 
in Figure~\ref{fig:simulation}a. 
After setting $m_x=d_x$, 
we specify $a_x$, $a_y$, $d_x$, and $d_y$,
and confirm an approximately $\pi$ phase difference
between on and off states for $y$ polarization.
After this, we consider an incidence of an $x$-polarized wave,
and adjust $m_x$ and $m_y$ to produce a $-\pi/2$ phase difference
between $x$- and $y$-polarized waves for the off state.
The parameters are chosen so that the helicity conversion
efficiency is as high and the working bandwidth
as broad as possible, while keeping feasible $d_y$:
they are determined 
to be $a_x=115 \,\mathrm{\mu m}$, $a_y=30 \,\mathrm{\mu m}$,
$a_z=26\,\mathrm{\mu m}$, 
$d_x=20 \,\mathrm{\mu m}$, $d_y=6 \,\mathrm{\mu m}$, 
$m_x=44\,\mathrm{\mu m}$, and $m_y=10 \,\mathrm{\mu m}$.

Figure~\ref{fig:simulation}b shows the calculated helicity-conversion
efficiency $S_3/S_\mathrm{in} = -\Im \tilde{r}_{x}^*\tilde{r}_y$,
along with the efficiency limit of a single-layer transmissive metasurface.
(Note that the propagation-direction change in reflection demands
the minus sign, in contrast to Equation~(\ref{eq:2}).)
At a frequency of $0.85\,\mathrm{THz}$, 
helicity conversion efficiency reaches 
$-0.94$ and $0.92$ for the off and on states, respectively. The degradation from perfect efficiency is caused by Ohmic and dielectric loss.
Nevertheless, the efficiencies of both off and on states significantly exceed 
the theoretical efficiency limit ($\approx 0.65$)
of the single-layer transmissive metasurface.
Detailed simulation results on the helicity conversion efficiency and
circular-polarization ratio for the intermediate state of vanadium dioxide 
are presented in Supplementary Information.

\subsection{Physical device}

The proposed metasurface design poses a fabrication challenge
because it involves multiple layers.
We prepared a thin sapphire plate with a metal ground
by room-temperature bonding with thin metal films \cite{Shimatsu2010}.
Because room-temperature bonding is free from glue,
it is tolerant of the high temperatures and chemical treatments required  
in the vanadium dioxide process.
The vanadium dioxide and aluminum patterns on the substrate were
fabricated by a conventional photolithography technique \cite{Nakata2019c,Nakanishi2020}.
Detailed information about the process is provided in the Experimental Section.

\begin{figure}
  \includegraphics{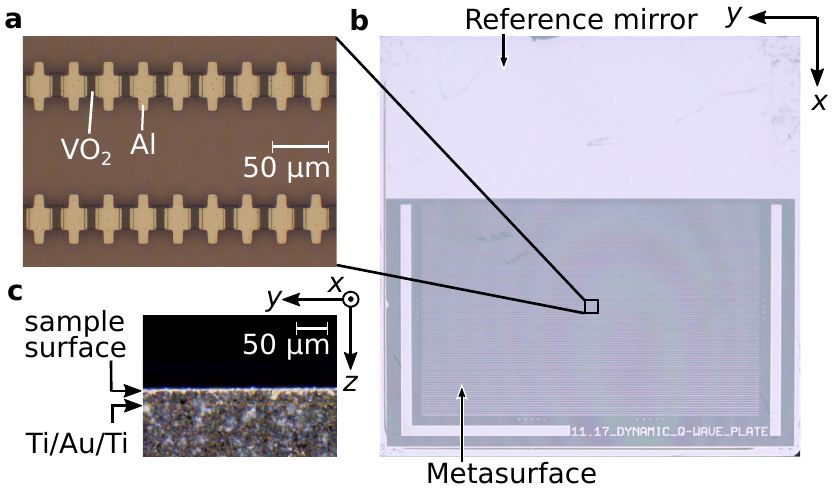}
  \caption{  a) Photomicrograph of vanadium dioxide (VO$_2$) and aluminum (Al) patterns fabricated on a substrate. 
b) Photograph of the entire sample. The upper half of the sample was covered with aluminum for use as a reference mirror.
c) Photomicrograph of the substrate viewed from the side. The sapphire plates are 
bonded at Ti/Au/Ti layers.
}
  \label{fig:sample}
\end{figure}

\textbf{Figure~\ref{fig:sample}}a and b show photomicrographs of the metasurface and a picture of the entire fabricated device, respectively.
Although there was an unintended shift of the 
vanadium-dioxide structures
relative to the aluminum ones, they connected with each other well.
The structural parameters measured for the actual sample are provided in the Experimental Section.
Figure~\ref{fig:sample}c offers a side view of the sample.
The sapphire plates were well-attached by the room-temperature bonding technique,
and the bonding was not broken after
high-temperature treatment and chemical etching.

To demonstrate the performance of the device, we conducted conventional terahertz time-domain spectroscopy.
Detailed information about this experiment is provided in the Experimental Section.
For comparison, we recalculate the device’s performance using the measured parameter values from the Experimental Section.
\textbf{Figure~\ref{fig:experimental_results}}a compares the helicity-conversion
efficiency $S_3/S_\mathrm{in}$ in the experiment and the
realistic simulation.
The experimental data agree well with the simulated ones.
At $f_0=0.90\,\mathrm{THz}$, the experimental helicity-conversion efficiency
reaches $-0.80$ and $0.87$ for the off and on states, respectively.
These are both far beyond $0.65$, the theoretical efficiency limit of single-layer transmissive metasurfaces.
To evaluate the improvement quantitatively, we compare the conversion efficiency with that of previous transmissive devices \cite{Nakata2019c,Nakanishi2020}. 
The helicity-conversion efficiency 
of the dipole-nested-checkerboard metasurface is 
$0.21$ and $-0.13$ at $0.617\,\mathrm{THz}$ \cite{Nakata2019c}, 
while that of the deformed-checkerboard metasurface is
$0.19$ and $-0.17$ at $0.66\,\mathrm{THz}$ \cite{Nakanishi2020}.
These efficiencies are much lower than the theoretical conversion efficiency of $0.65$,
because of the multiple reflections inside the substrates, and
the efficiencies are evaluated for the first transmitted pulse.
The efficiency of the proposed device is more than four times higher,
because it does not suffer from the scattering loss from the substrate and metasurface.
Other reported devices do not achieve orthogonal modulation,
and their transmission efficiencies are similar to the above checkerboard devices
because of substrate reflections \cite{Zhao2018a, Wang2015b, Wang2016, Nouman2018a}.

We also evaluated the circular-polarization ratio $S_3/S_\mathrm{out}$ to determine the purity of the output polarization.
Figure~\ref{fig:experimental_results}b shows experimental and simulated $S_3/S_\mathrm{out}$.
The experimental circular-polarization ratio $S_3/S_\mathrm{out}$ at
$f_0=0.90\,\mathrm{THz}$
is $-0.96$ for the off state and $1.00$ for the on state.
These data indicate that each reflected wave is in a state close to pure circular polarization.
Thus, our device is not only highly efficient, but also achieves deep modulation between orthogonal circular polarizations.

The bandwidth of the device ($|S_3/S_\mathrm{out}|>0.5$ for both the ON and OFF states)
is estimated as $\Delta f = 0.22\,\mathrm{THz}$.
The relative bandwidth 
is given by $\Delta f/f_0 = 0.25$,
which is wider than that of the dipole-nested checkerboard metasurface
\cite{Nakata2019c} ($\Delta f/f_0 = 0.12 $),
but narrower than that of the deformed checkerboard \cite{Nakanishi2020} ($\Delta f/f_0 = 0.52$).

\subsection{Discussion}

To show the details of the device operation, 
we summarize the reflection amplitudes $|\tilde{r}_x|$ and $|\tilde{r}_y|$
and the phase difference $\arg(\tilde{r}_y)-\arg(\tilde{r}_x)$
in \textbf{Figure~\ref{fig:detailed_experimental_results}}.
The solid lines represent the measured data, and the dashed lines represent the simulated data; the agreement of the two is generally good.
The data confirm remarkably high reflectivity, which leads to a high device efficiency up to $1.0\,\mathrm{THz}$. 
The $\pi/2$ phase shift between $x$ and $y$ polarizations 
is realized at $f_0=0.90\,\mathrm{THz}$.
Interestingly, the phase difference of the off state 
showed a plateau around $f_0$, which indicates
that the phase rotations of $\tilde{r}_x$ and $\tilde{r}_y$ with increasing $\omega$
cancelled out.
However, such a phase plateau was not observed for the on state.
If an on-state plateau could be realized, we could achieve a much broader operation.
The $|\tilde{r}_x|$ dips around $1.1\,\mathrm{THz}$
can be explained by energy leakage to the waveguide modes.
Because the structures act as grating couplers, the incident wave can couple to the waveguide mode.
The frequency of the transverse magnetic (TM) waveguide mode with grating wavenumber $G_x=2\pi /a_x$
is theoretically estimated to be $1.1\,\mathrm{THz}$ \cite{Pozar2011},
assuming that sapphire has an isotropic refractive index of $3.4$.

The stability of the device is also important for applications.
In Supplemental Information, we present the reflection characteristics of the device measured at different cycles of heating and cooling operations on three different days. No performance degradation was observed in the cycles.
Moreover, a previous report indicates that hundred cycles of operations do not alter the VO$_2$ characteristics \cite{Ko2008}.

The underlying physics can be elucidated by an eigenmode analysis of the realistic model without excitation.
As impedance and transition boundaries cannot be used in eigenmode analysis,
we represent the aluminum and gold regions as  
perfect electric conductors in this simulation.
The calculated eigenfrequencies are depicted as gray lines in
Figure~\ref{fig:detailed_experimental_results}.
The corresponding eigenmodes, depicted in \textbf{Figure~\ref{fig:eigenmodes}}, induce phase rotations around the resonant frequencies.
Eigenmodes with current flowing along the $x$ and $y$ axes are termed $x$- and $y$-polarized modes, respectively.
For the off state, the $x$-polarized mode at $0.76\,\mathrm{THz}$ [Figure~\ref{fig:eigenmodes}a] and the
$y$-polarized mode at $0.87\,\mathrm{THz}$ [Figure~\ref{fig:eigenmodes}d]
maintain $\arg(\tilde{r}_y)-\arg(\tilde{r}_x) \approx \pi/2$ around the
working frequency of $0.90\,\mathrm{THz}$.
The $y$-polarized modes at $0.87\,\mathrm{THz}$ and $1.40\,\mathrm{THz}$ in the off state 
merge into another resonant mode at $1.22\,\mathrm{THz}$ in the on state.
This change causes a $\pi$-radian phase shift for target $y$ polarization.
The modes around $1.10\,\mathrm{THz}$ and $1.15\,\mathrm{THz}$ are split from the waveguide modes by introducing a periodic structure.
As their currents flow in the $x$ direction, 
their frequencies do not shift with switching from off to on. 
The waveguide modes at $1.15\,\mathrm{THz}$ have antisymmetric current distributions 
in the $x$ direction.
The antisymmetric waveguide mode cannot be excited in the simulation because of the symmetry of the incident plane wave. However, it can be excited experimentally,
because of the spatially-nonuniform field distribution of the incident wave.
In practice, we used a lens to focus the incident wave, as explained in the Experimental section.
This antisymmetric waveguide mode explains the reflection dip at $1.15\,\mathrm{THz}$,
which was only observed in experiments.
The excited waveguide mode propagates 
inside the sapphire and can emit from the side.
This leakage might contribute to the linewidth broadening of the dips
around $1.15\,\mathrm{THz}$, as shown in Figure~\ref{fig:detailed_experimental_results}a and b.

\section{Conclusion}

In this study, we achieved orthogonal-polarization modulation with high conversion efficiency
in terahertz helicity switching by leveraging a dynamic reflective
quarter-wave metasurface.
The helicity-conversion efficiency reached $-80\,\%$ and
$87\,\%$ at $0.90\,\mathrm{THz}$, 
which is far beyond the efficiency limit 
for single-layer transmissive metasurfaces ($65\,\%$).
Our device had more than four times the efficiency of devices in previous studies \cite{Nakata2019c,Nakanishi2020}.
In addition, the purity of its output helicity reached $-96\,\%$
and $100\,\%$.
These results indicate the decisive advantage of using dynamic reflective
metasurfaces for orthogonal-polarization modulation.

Compared to PIN-diode approach for a microwave helicity switchable metasurface
under circular-polarization illumination \cite{Xu2016a},
vanadium-dioxide devices are free from soldering
and allow operation in terahertz frequency range.
In addition, our device works as a quarter-wave plate 
that does not require circular-polarization illumination for helicity switching.
Due to the scale invariance of Maxwell's equations, 
the proposed design scheme is applicable not only in the terahertz frequency
range, but also in other frequency ranges from microwave to infrared regions,
using suitable metal, dielectric, or high-contrast
conductivity-changeable materials.
To reach the higher frequency range, 
GST (Ge$_2$Sb$_2$Te$_5$) might be a suitable choice of
dielectric phase-changeable material \cite{Zhang2021}.
Although the switching speed of our device is limited by 
the heater response time ($\sim$ minutes) in our experiments,
the potential switching speed of vanadium dioxide is much faster \cite{Mian2015}.
Ultrafast switching could be realized by photoexcitation of
vanadium dioxide \cite{Kim2019} or semiconductors \cite{Yang2017}.
Moreover, broadening of the operating bandwidth of the device is another interesting problem.
Checkerboard structures might be helpful for achieving broadband
operation \cite{Nakanishi2020}.
The room-temperature-bonding technique developed here
is compatible with high temperature and chemical etching,
and is applicable to switchable multifunctional terahertz devices \cite{Liang2021, Shabanpour2020, Ren2021a}.


\section{Experimental Section}
\threesubsection{Substrate Bonding}\\

Two substrates of c-plane sapphire 
were prepared and coated with titanium (thickness: $5\,\mathrm{nm}$) on
their top surfaces as an adhesion layer. A layer of gold 
$200\,\mathrm{nm}$ thick
was deposited on the titanium on one sample,
while a gold layer $20\,\mathrm{nm}$ thick was formed on the titanium on the other sample.
The extremely flat gold sides of these substrates were bonded by gold diffusion at room temperature: this worked like glue, 
without introducing other materials. 
This room-temperature
bonding technique is highly compatible with the vanadium-dioxide process,
which involves high-temperature and chemical treatment.
Next, one side of the sapphire was 
polished to a thickness of $26\,\mathrm{\mu m}$.
Finally, the substrate was cut into 
$20\,\mathrm{mm}\times20\,\mathrm{mm}\times 1 \,\mathrm{mm}$ plates.

\threesubsection{Photolithography}\\

First, we deposited a $250\,\mathrm{nm}$-thick VO$_2$
film on the bonded sapphire substrate.
We used reactive magnetron sputtering with a vanadium 
target in O$_2$ and Ar flow. During the sputtering, 
the temperature of the substrate was kept at 600\,$^\circ\mathrm{C}$.
Next, we coated the photoresist on the VO$_2$ film
and exposed it, using a maskless exposure system.
After the development of the photoresist, the VO$_2$ pattern was formed
by wet etching in a solution comprising
phosphoric acid, nitric acid, acetic acid, and water in the proportions 16:1:2:1.
Because the vanadium-dioxide patterns were over-etched,
the exposure was intentionally $2 \,\mathrm{\mu m}$ larger than the original design parameter.
The aluminum pattern was fabricated by lift-off techniques. 
Half of the surface had the structures, 
while the other half was a uniform aluminum sheet used as a reference mirror. 
By coating, exposing, and developing the photoresist,
we left it in place except in the pattern region.
Finally, a $400\,\mathrm{nm}$-thick aluminum pattern was formed
by electron-beam evaporation and lift off.

\threesubsection{Sample-size characterization}\\

\begin{figure}
  \includegraphics{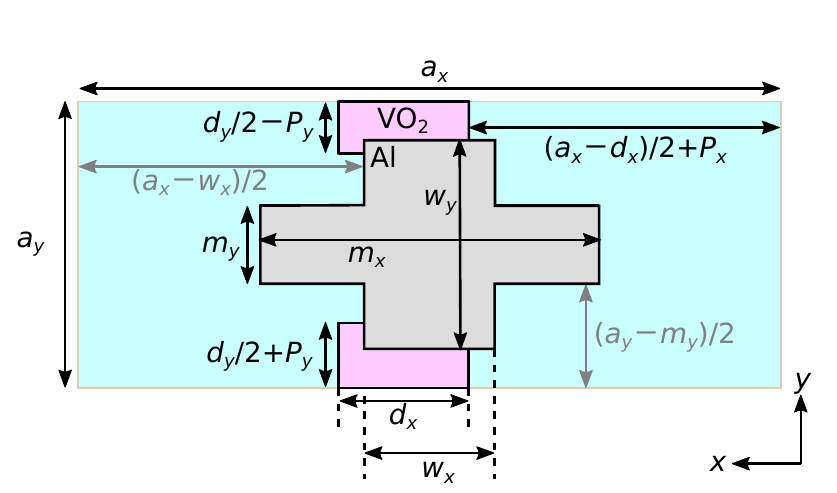}
  \caption{Dimensional parameters of a realistic unit cell of the device. The offset of VO$_2$ from its original position is represented by $P_x$ and $P_y$; the Al structure is placed at the center.
  \label{fig:realistic_simulation_model}}
\end{figure}

The size of the fabricated sample was characterized by optical microscopy.
The geometric parameters are defined in \textbf{Figure~\ref{fig:realistic_simulation_model}}.
Their measured values were as follows:

$a_x=115 \,\mathrm{\mu m}$, $a_y=30 \,\mathrm{\mu m}$,
$d_x=19.9 \,\mathrm{\mu m}$, $d_y=11.7 \,\mathrm{\mu m}$,
$w_x=19.5 \,\mathrm{\mu m}$, $w_y=23.0 \,\mathrm{\mu m}$,  
$m_x=42.6\,\mathrm{\mu m}$, and $m_y=8.9 \,\mathrm{\mu m}$.
The displacement of the VO$_2$ pattern was evaluated as
$P_x=-1.6\,\mathrm{\mu m}$ and $P_y=-0.35\,\mathrm{\mu m}$.
The thickness of the substrate was estimated to be
$a_z=24.4\,\mathrm{\mu m}$ by observing 
interference fringes of the reflected white light with a spectrometer,
assuming the refractive index of sapphire to be 1.76.
The direct-current (DC) sheet resistance of the vanadium dioxide was
measured as $1\times 10^2\,\mathrm{k}\Omega/\square$ in the off state ($300\,\mathrm{K}$) 
and $10 \,\Omega/\square$ in the on state ($370\, \mathrm{K}$).

\begin{figure}
  \includegraphics{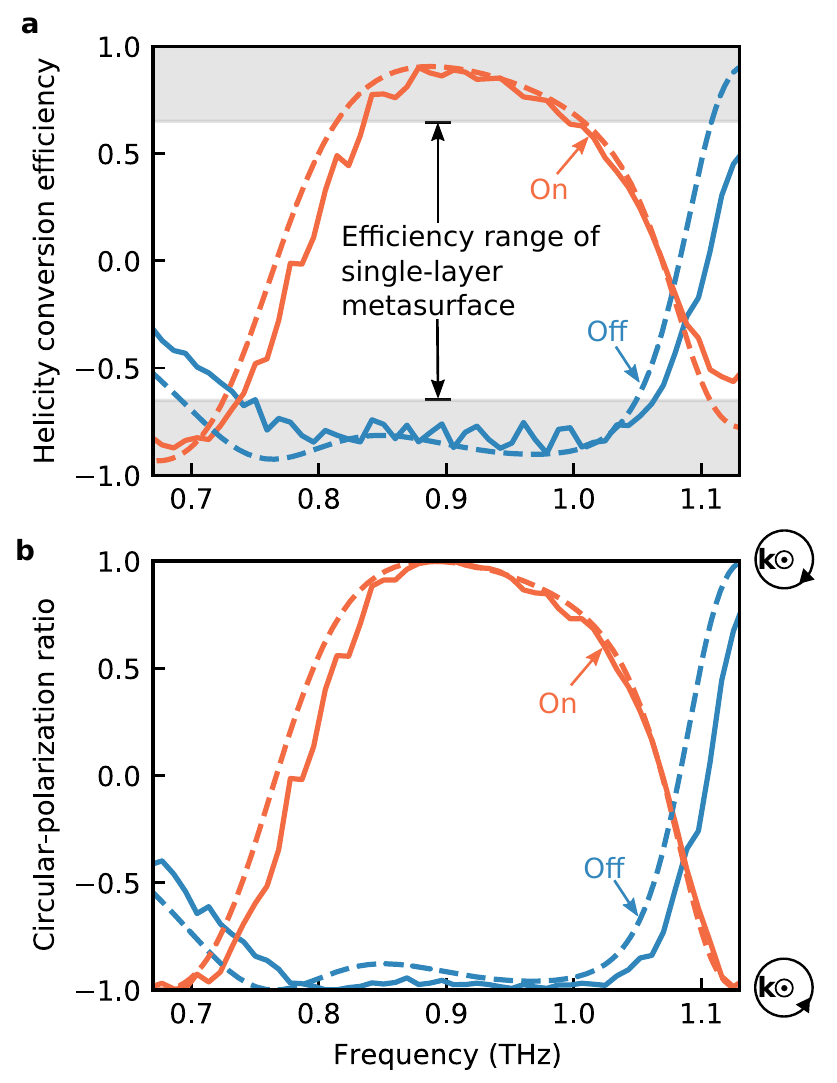}
  \caption{ Measured and simulated data for a) helicity-conversion efficiency $S_3/S_{\mathrm{in}} $ and b) circular-polarization ratio $S_3/S_{\mathrm{out}}$. 
Solid and dashed lines represent measured and simulated data, respectively.
The parameters shown in the Experimental Section are used in this simulation.
}
  \label{fig:experimental_results}
\end{figure}

\begin{figure}
  \includegraphics{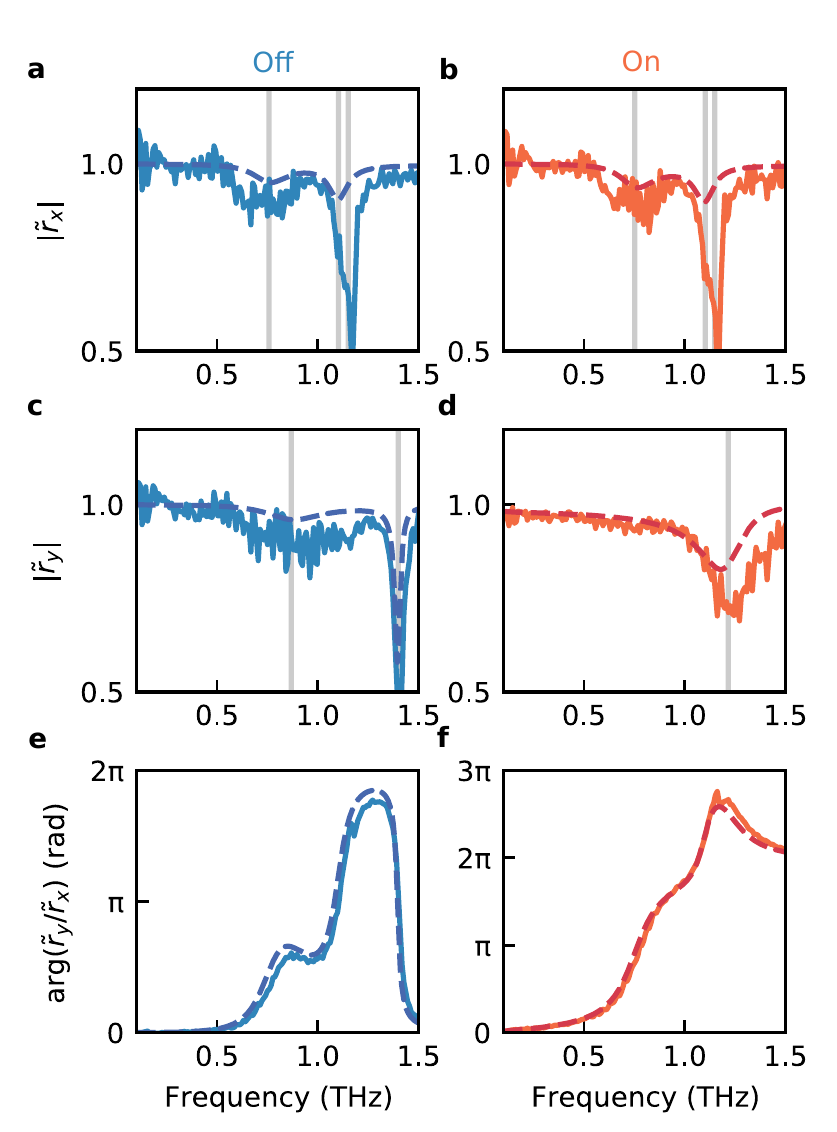}
  \caption{
Frequency dependence of reflection amplitudes of 
$x$ polarization for a) off and b) on states,
those of $y$ polarization for c) off and d) on states,
and of phase difference between $x$ and $y$ polarizations for e) off and f) on states.
Solid and dashed lines represent the measured and simulated data, respectively.
The gray vertical lines indicate the eigenfrequencies of the structure.
The parameters shown in the Experimental Section are used in this simulation.
}
  \label{fig:detailed_experimental_results}
\end{figure}

\begin{figure}
  \includegraphics{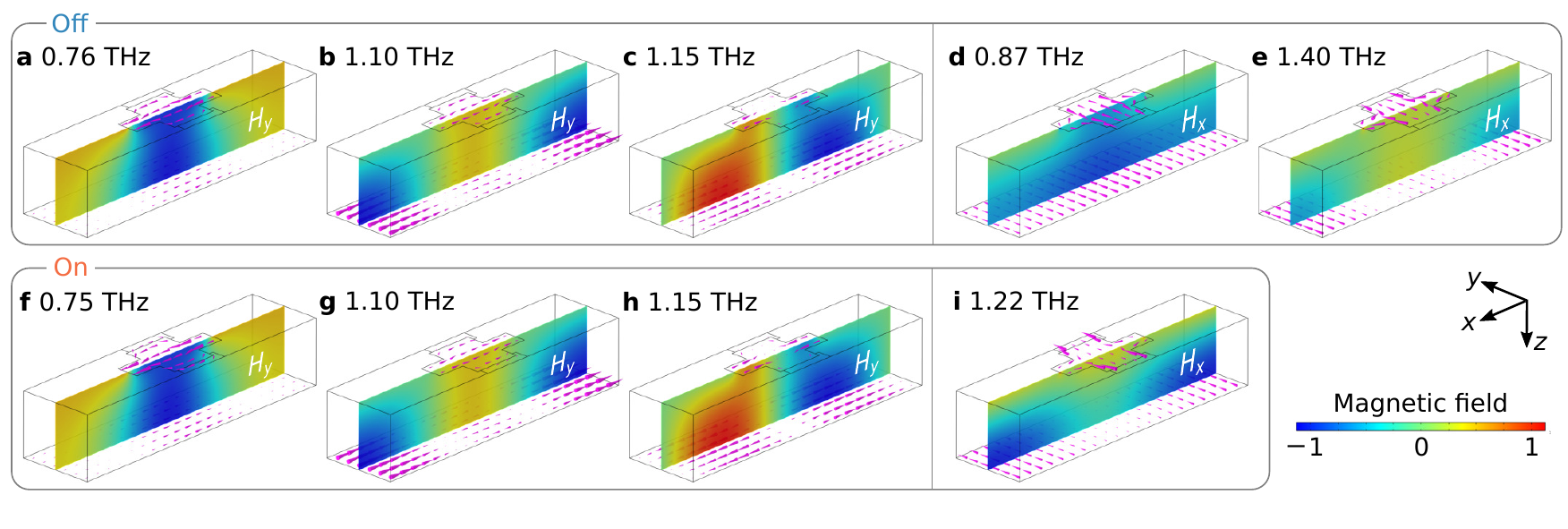}
  \caption{Current and magnetic-field distributions of eigenmodes in the realistic model. a)--e) Off-state. f)--i) On state. 
The current is represented by magenta arrows, while color-map plots
express the magnetic fields $H_y$ [a)--c) and f)--h)] of $x$-polarized modes 
and $H_x$ [d), e), and i)] of the $y$-polarized modes.
The parameters shown in the Experimental Section were used in the calculation.
}
  \label{fig:eigenmodes}
\end{figure}

These parameters were used in the realistic simulation discussed in Figure~\ref{fig:experimental_results}, \ref{fig:detailed_experimental_results}, and \ref{fig:eigenmodes}.
When calculating reflection spectra, 
the boundary conditions used for aluminum and bottom gold were kept
unchanged from the design simulation.
The aluminum and gold were assumed to be perfect conductors in the eigenmode analysis.

\threesubsection{Terahertz Reflection Measurement}\\

\begin{figure}
  \includegraphics{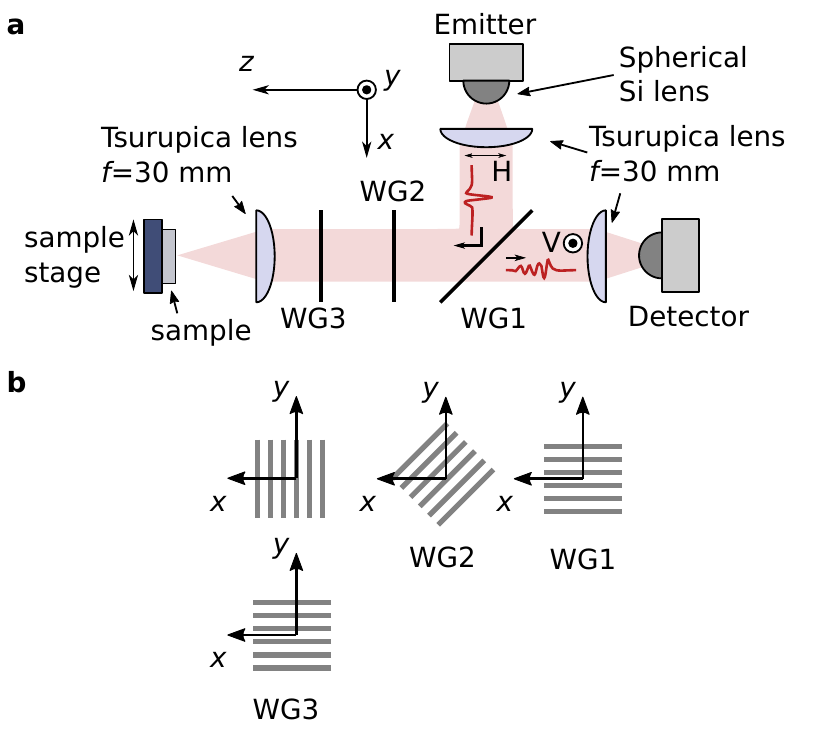}
  \caption{a) Experimental setup of reflection measurement for $x$ and $y$ polarizations. H and V indicate horizontal and vertical polarizations.
b) Wire-grid (WG) configurations. The wires of WG3 are set parallel to $y$ or $x$ in $\tilde{r}_x$ or $\tilde{r}_y$ complex-reflectivity measurements, respectively.}
  \label{fig:exp_setup}
\end{figure}

Conventional terahertz time-domain spectroscopy was performed in a reflective setup.
\textbf{Figure~\ref{fig:exp_setup}} shows a schematic picture of the terahertz system.
Photoconductive antennas were excited by femtosecond-laser pulses
to emit and detect a terahertz pulse in the time domain.
Horizontally polarized radiation from 
the emitter was collimated by a lens and then reflected by the first wire grid (WG1). 
As the wires of WG1 were parallel to the $x$ axis,
the incident wave was mostly reflected by them and then passed through grids WG2 and WG3 before reaching the sample. 
The wires in WG2 were aligned along the diagonal ($y=-x$) direction, while those in WG3 were set
parallel to the $y$ or $x$ directions for the $\tilde{r}_x$ or $\tilde{r}_y$ measurements, respectively.
Eventually, the wave was focused on the sample by a lens.
The full width at half maximum (FWHM) of the beam spot was $2.5\,\mathrm{mm}$.
Radiation reflected by the sample then partially passed through WG2.
Finally, a part of the wave was transmitted through WG1
to the detector.
The measurements were taken in a sealed box filled with dry air.
First, we measured the temporal reference signal $E_i^{(\mathrm{ref})}(t)$ reflected from the aluminum reference mirror on the sample.
After translating the sample holder, we obtained the temporal signal $E_i(t)$ reflected from the metasurface.
To ensure phase accuracy,
the displacement direction of the sample stage was 
precisely adjusted to be parallel to the sample surface.
We took the Fourier transform of the signals and calculated the complex reflection amplitudes 
$\tilde{r}_i(\omega)=\tilde{E}_i(\omega)/\tilde{E}_i^{(\mathrm{ref})}(\omega)$.
We averaged the reflection-amplitude data obtained by repeating this procedure ten times.
The measurement was performed for four sets of $(i,T)$, where $i=x,y$ and 
metasurface-temperature $T=300$ (off) or $370\,\mathrm{K}$ (on).


\medskip
\textbf{Acknowledgements} \par 

The authors would like to thank Hitoe Kon in Adamant Namiki Precision Jewel Co., Ltd. for her technical contribution to sapphire bonding.
We are also grateful to Akira Inokuma and Yuma Takano for their helpful comments on the manuscript, and Kei Iyoda and Masayuki Fujita for technical assistance. 
This study was supported by a grant from the Murata Science Foundation.
A part of this work was supported by the Kyoto University Nano Technology Hub in “Nanotechnology Platform Project” sponsored by the Ministry of Education, Culture, Sports, Science and Technology (MEXT), Japan.

\medskip

Dynamic polarization control is essential for sensing and communication.
A dynamic metasurface that uses phase-changeable vanadium dioxide to modulate between circular polarizations is described. It has a multilayer structure, fabricated using a room-temperature bonding method.
Almost perfect helicity inversion is demonstrated in the terahertz range;
the conversion efficiency ($80\,\%$)
breaks the fundamental efficiency limit of single-layer metasurfaces.


\begin{figure}
\textbf{Table of Contents}\\
\medskip
\includegraphics{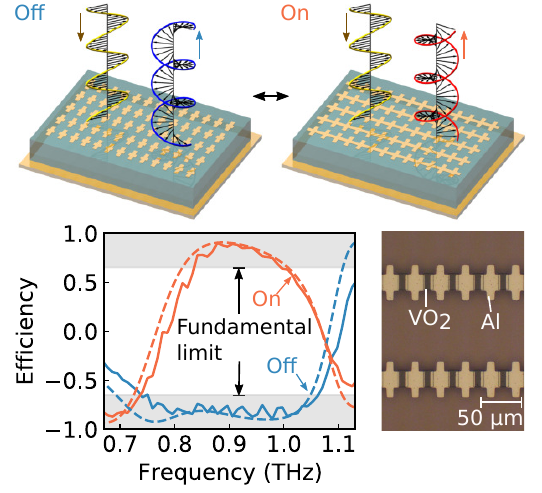}
  \medskip
  \caption*{ToC Entry}
\end{figure}

\graphicspath{{./figs/}{./figs/SI}}  

\def\Re{\mathop{\mathrm{Re}}}
\def\Im{\mathop{\mathrm{Im}}}

\renewcommand{\thefigure}{S\arabic{figure}}

\clearpage
\setcounter{equation}{0}
\setcounter{figure}{0}
\setcounter{table}{0}
\setcounter{page}{1}
\makeatletter
\renewcommand{\theequation}{S\arabic{equation}}
\renewcommand{\thefigure}{S\arabic{figure}}

\renewcommand{\theHtable}{Supplement.\thetable}
\renewcommand{\theHfigure}{Supplement.\thefigure}
\renewcommand{\theHequation}{Supplement.\theequation}

\title{
Supporting Information 
for ``Dynamic quarter-wave metasurface for efficient helicity inversion
of polarization beyond the single-layer conversion limit''
}
\maketitle

\author{Mitsuki Kobachi}
\author{Fumiaki Miyamaru}
\author{Toshihiro Nakanishi}
\author{Kunio Okimura}
\author{Atsushi Sanada}

\author{Yosuke Nakata*}







\dedication{}

\section{Calculated performance of the design in intermediate states}

\begin{figure}[htbp]
 \begin{center}
 \includegraphics{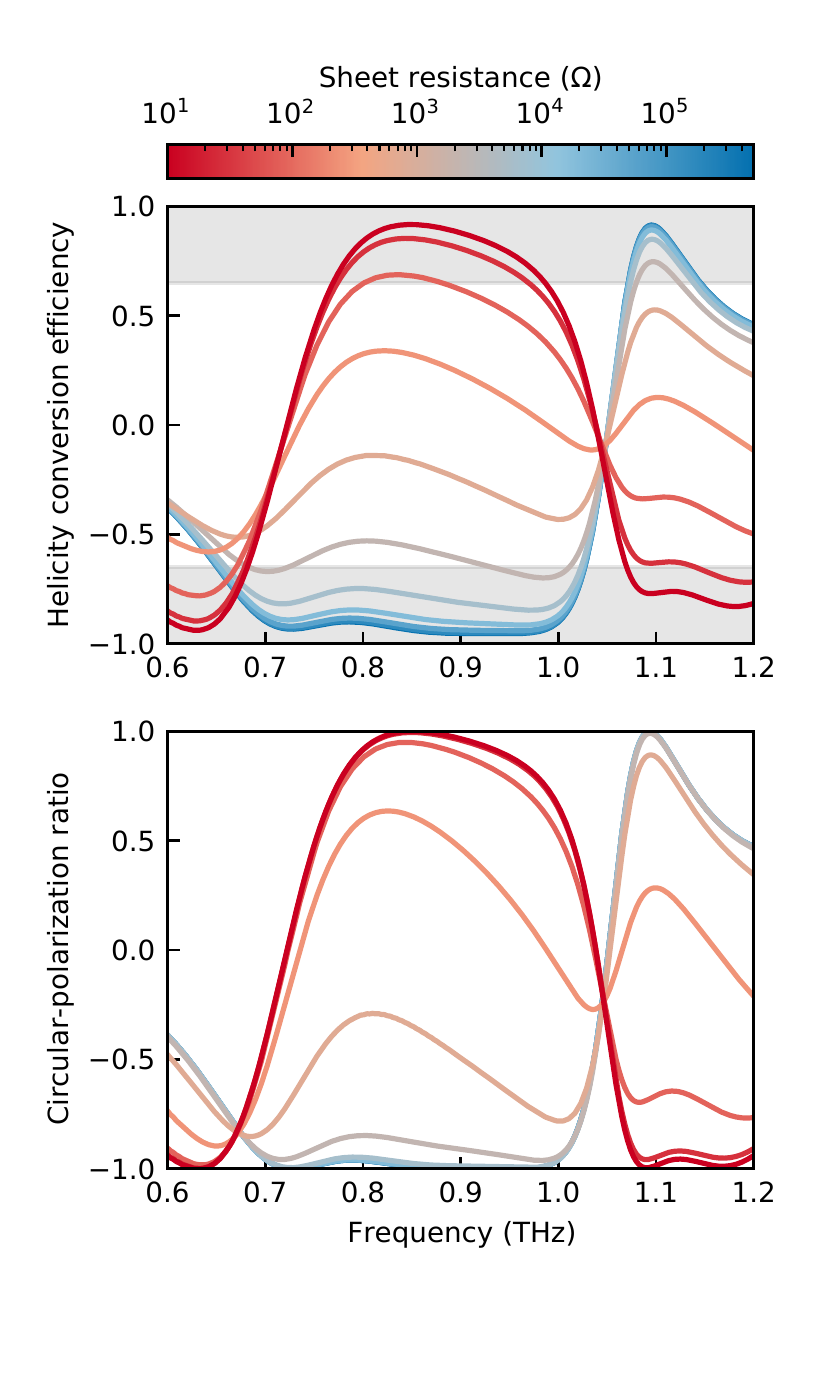}
 \end{center}  
  \caption{Helicity conversion efficiency and circular polarization ratio when varying the value of VO$_2$ sheet resistance between 
10\,$\Omega$ to 500\,$\mathrm{k}\Omega$ with a logarithmic scale
in the simulation.
The calculation setup is the same as that in Figure~2 in the manuscript.
}
  \label{fig:stokes}
\end{figure}

\begin{figure}[htbp]
 \begin{center}
 \includegraphics{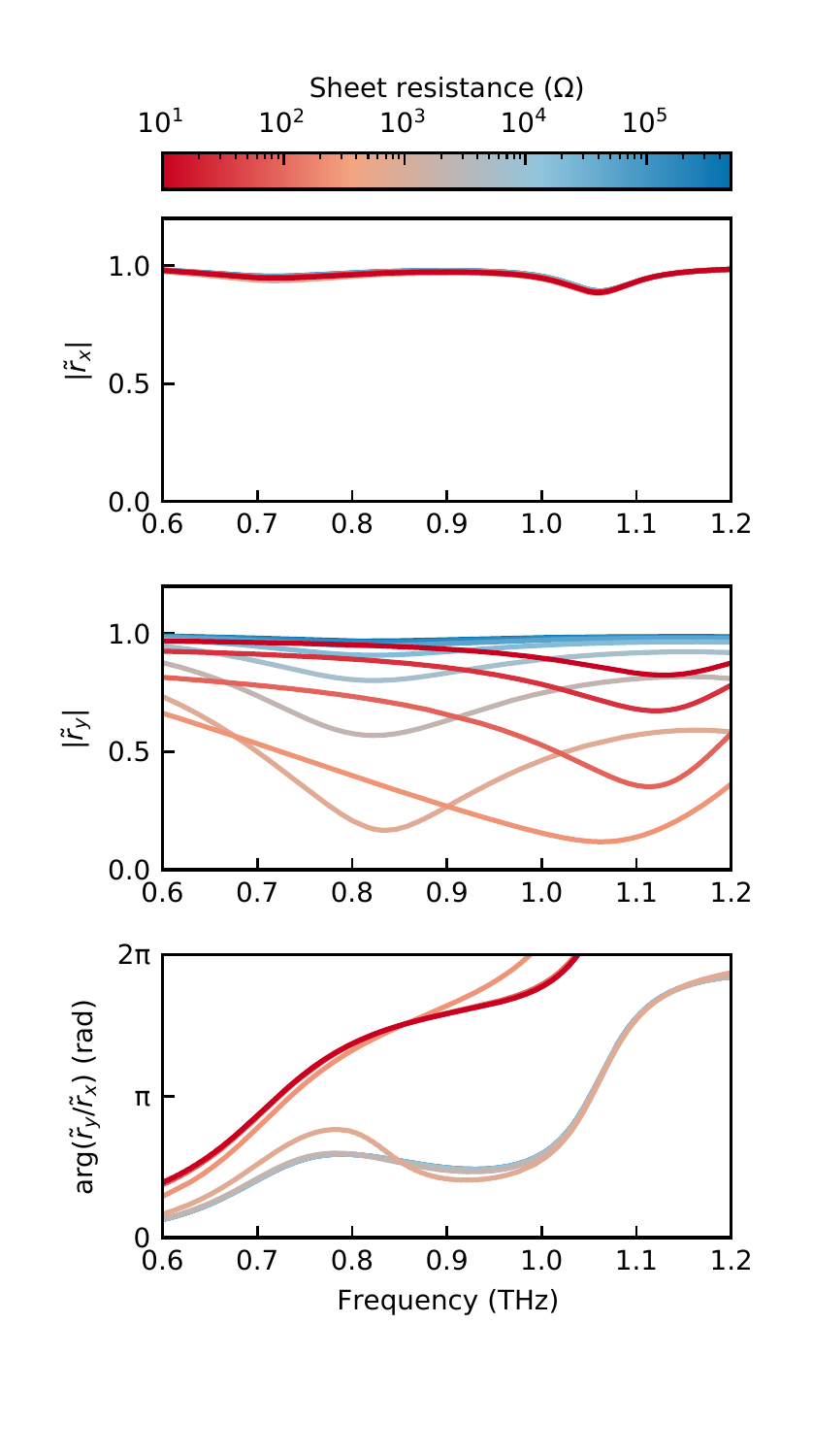}
 \end{center}  
  \caption{Amplitude and phase difference of $x$ and $y$ polarizations of reflected waves when the value of VO$_2$ sheet resistance is varied between 
10\,$\Omega$ to 500\,$\mathrm{k}\Omega$ with a logarithmic scale in the simulation.
The calculation setup is the same as that in Figure~2 in the manuscript.
}
  \label{fig:amp_phase}
\end{figure}

Figure~\ref{fig:stokes} represents the VO$_2$-resistance 
dependency of the helicity conversion efficiency and circular polarization ratio for the original design.
The gradual transition between the off and on states can be observed.
Figure~\ref{fig:amp_phase} shows the corresponding reflection amplitude and phase.
The reflection $|\tilde{r}_x|$ remains unchanged as the current barely flows in VO$_2$.
On the other hand, $|\tilde{r}_y|$ 
has dissipation in the intermediate states
due to ohmic loss.
The transition of the resonant frequency induces a phase jump in $\arg(\tilde{r}_y/\tilde{r}_x)$.

\clearpage

\section{Stability of the fabricated device}

\begin{figure}[htbp]
 \begin{center}
 \includegraphics{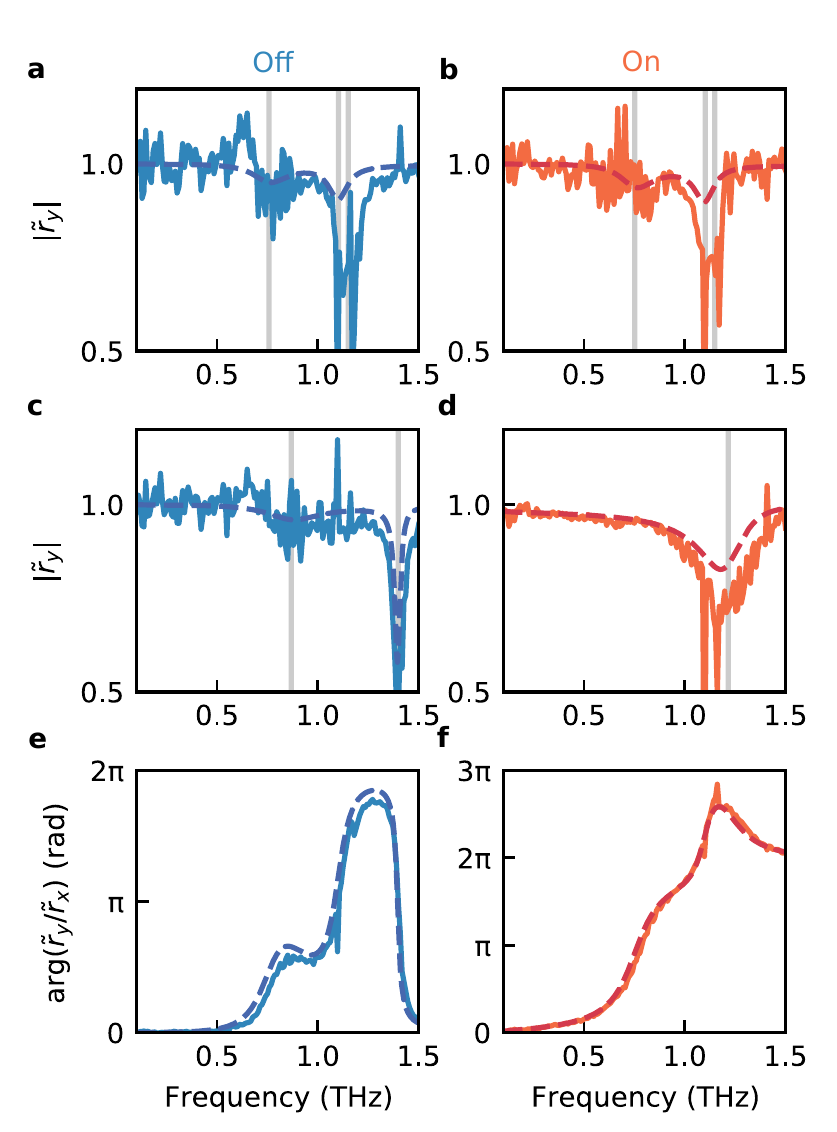}
 \end{center}  
  \caption{Amplitude and phase difference for $x$ and $y$ polarizations based on data measured on March 16, 2021, without dry air.
The solid and dashed lines represent the experimental and simulation data, respectively. The gray vertical lines indicate the eigenfrequencies
of the structure. The parameters shown in the Experimental section are used in this simulation. 
}
  \label{fig:amp_phase_210316}
\end{figure}

\begin{figure}[htbp]
 \begin{center}
 \includegraphics{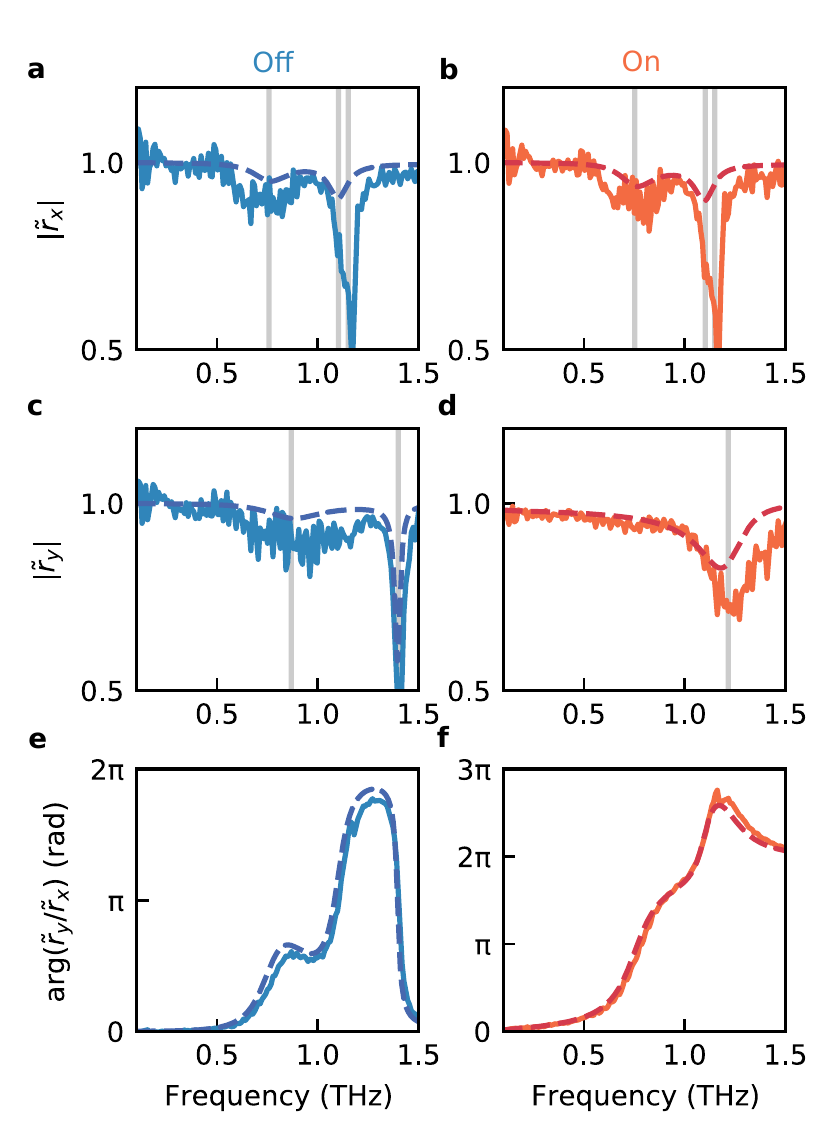}
 \end{center}  
  \caption{Amplitude and phase difference for $x$ and $y$ polarizations based on data measured on March 17, 2021 (used in the manuscript).
The solid and dashed lines represent the experimental and simulation data, respectively. The gray vertical lines indicate the eigenfrequencies
of the structure. The parameters shown in the Experimental section are used in this simulation.
}
  \label{fig:amp_phase_210317}
\end{figure}

\begin{figure}[htbp]
 \begin{center}
 \includegraphics{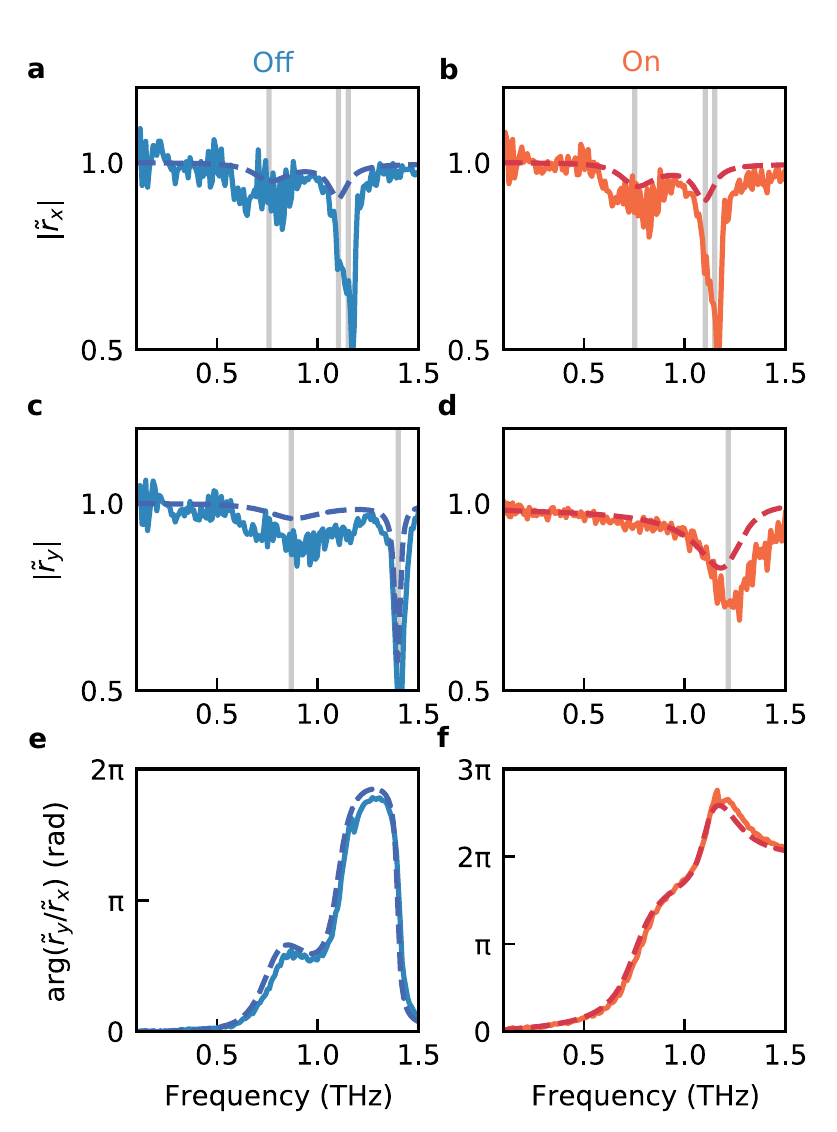}
 \end{center}  
  \caption{Amplitude and phase difference for $x$ and $y$ polarizations based on data measured on March 18, 2021.
The solid and dashed lines represent the experimental and simulation data, respectively. The gray vertical lines indicate the eigenfrequencies
of the structure. The parameters shown in the Experimental section are used in this simulation.
}
  \label{fig:amp_phase_210318}
\end{figure}

Figures~\ref{fig:amp_phase_210316}, \ref{fig:amp_phase_210317} (used in the manuscript), and \ref{fig:amp_phase_210318} show the device characteristics measured at different cycles on three different days.
These data certainly confirm the reproducibility of our device.
No performance degradation is observed in the cycles.
Note that we did not fill the sample box with dry air for the data shown in Figure~\ref{fig:amp_phase_210316}; thus, the data show dips of terahertz absorption by water.

\begin{thebibliography}{10}
\providecommand{\url}[1]{\texttt{#1}}
\providecommand{\urlprefix}{URL }

\bibitem{Chen2016}
H.-T. Chen, A.~J. Taylor, N.~Yu,
\newblock \emph{Rep. Prog. Phys.} \textbf{2016}, \emph{79}, 7 076401.

\bibitem{He2018}
Q.~He, S.~Sun, S.~Xiao, L.~Zhou,
\newblock \emph{Adv. Opt. Mater.} \textbf{2018}, \emph{6}, 19 1800415.

\bibitem{Takano2011}
K.~Takano, H.~Yokoyama, A.~Ichii, I.~Morimoto, M.~Hangyo,
\newblock \emph{Opt. Lett.} \textbf{2011}, \emph{36}, 14 2665.

\bibitem{Suzuki2016}
T.~Suzuki, M.~Nagai, Y.~Kishi,
\newblock \emph{Opt. Lett.} \textbf{2016}, \emph{41}, 2 325.

\bibitem{Grady2013}
N.~K. Grady, J.~E. Heyes, D.~R. Chowdhury, Y.~Zeng, M.~T. Reiten, A.~K. Azad,
  A.~J. Taylor, D.~A.~R. Dalvit, H.-T. Chen,
\newblock \emph{Science} \textbf{2013}, \emph{340}, 6138 1304.

\bibitem{Ding2015}
X.~Ding, F.~Monticone, K.~Zhang, L.~Zhang, D.~Gao, S.~N. Burokur,
  A.~de~Lustrac, Q.~Wu, C.-W. Qiu, A.~Al{\`{u}},
\newblock \emph{Adv. Mater.} \textbf{2015}, \emph{27}, 7 1195.

\bibitem{Nakata2017}
Y.~Nakata, Y.~Taira, T.~Nakanishi, F.~Miyamaru,
\newblock \emph{Opt. Express} \textbf{2017}, \emph{25}, 3 2107.

\bibitem{Han2018}
Z.~Han, S.~Ohno, Y.~Tokizane, K.~Nawata, T.~Notake, Y.~Takida, H.~Minamide,
\newblock \emph{Opt. Lett.} \textbf{2018}, \emph{43}, 12 2977.

\bibitem{Liu2019a}
W.~Liu, T.~Yu, Y.~Sun, Z.~Lai, Q.~Liao, T.~Wang, L.~Yu, H.~Chen,
\newblock \emph{Phys. Rev. Appl.} \textbf{2019}, \emph{11}, 6 064005.

\bibitem{You2020}
X.~You, R.~T. Ako, W.~S.~L. Lee, M.~Bhaskaran, S.~Sriram, C.~Fumeaux,
  W.~Withayachumnankul,
\newblock \emph{APL Photonics} \textbf{2020}, \emph{5}, 9 096108.

\bibitem{Herrmann2020}
E.~Herrmann, H.~Gao, Z.~Huang, S.~R. Sitaram, K.~Ma, X.~Wang,
\newblock \emph{J. Appl. Phys.} \textbf{2020}, \emph{128}, 14 140903.

\bibitem{Konishi2020}
K.~Konishi, T.~Kan, M.~Kuwata-Gonokami,
\newblock \emph{J. Appl. Phys.} \textbf{2020}, \emph{127}, 23 230902.

\bibitem{Zhu2013}
B.~O. Zhu, J.~Zhao, Y.~Feng,
\newblock \emph{Sci. Rep.} \textbf{2013}, \emph{3}, 1 3059.

\bibitem{Wu2019}
Z.~Wu, Y.~Ra'di, A.~Grbic,
\newblock \emph{Phys. Rev. X} \textbf{2019}, \emph{9}, 1 011036.

\bibitem{Kiani2020}
M.~Kiani, A.~Momeni, M.~Tayarani, C.~Ding,
\newblock \emph{Opt. Express} \textbf{2020}, \emph{28}, 23 35128.

\bibitem{Zhang2018}
Y.~Zhang, Y.~Zhao, S.~Liang, B.~Zhang, L.~Wang, T.~Zhou, W.~Kou, F.~Lan,
  H.~Zeng, J.~Han, Z.~Feng, Q.~Chen, P.~Mazumder, Z.~Yang,
\newblock \emph{Nanophotonics} \textbf{2018}, \emph{8}, 1 153.

\bibitem{Miao2015}
Z.~Miao, Q.~Wu, X.~Li, Q.~He, K.~Ding, Z.~An, Y.~Zhang, L.~Zhou,
\newblock \emph{Phys. Rev. X} \textbf{2015}, \emph{5}, 4 041027.

\bibitem{Okada2010}
T.~Okada, K.~Ooi, Y.~Nakata, K.~Fujita, K.~Tanaka, K.~Tanaka,
\newblock \emph{Opt. Lett.} \textbf{2010}, \emph{35}, 10 1719.

\bibitem{Yang2017}
Y.~Yang, N.~Kamaraju, S.~Campione, S.~Liu, J.~L. Reno, M.~B. Sinclair, R.~P.
  Prasankumar, I.~Brener,
\newblock \emph{ACS Photonics} \textbf{2017}, \emph{4}, 1 15.

\bibitem{Kanda2009}
N.~Kanda, K.~Konishi, M.~Kuwata-Gonokami,
\newblock \emph{Opt. Lett.} \textbf{2009}, \emph{34}, 19 3000.

\bibitem{Zhang2012}
S.~Zhang, J.~Zhou, Y.-S. Park, J.~Rho, R.~Singh, S.~Nam, A.~K. Azad, H.-T.
  Chen, X.~Yin, A.~J. Taylor, X.~Zhang,
\newblock \emph{Nat. Commun.} \textbf{2012}, \emph{3} 942.

\bibitem{Kanda2014}
N.~Kanda, K.~Konishi, M.~Kuwata-Gonokami,
\newblock \emph{Opt. Lett.} \textbf{2014}, \emph{39}, 11 3274.

\bibitem{Kan2013}
T.~Kan, A.~Isozaki, N.~Kanda, N.~Nemoto, K.~Konishi, M.~Kuwata-Gonokami,
  K.~Matsumoto, I.~Shimoyama,
\newblock \emph{Appl. Phys. Lett.} \textbf{2013}, \emph{102}, 22 221906.

\bibitem{Kan2015}
T.~Kan, A.~Isozaki, N.~Kanda, N.~Nemoto, K.~Konishi, H.~Takahashi,
  M.~Kuwata-Gonokami, K.~Matsumoto, I.~Shimoyama,
\newblock \emph{Nat. Commun.} \textbf{2015}, \emph{6} 8422.

\bibitem{Zhang2017}
M.~Zhang, W.~Zhang, A.~Q. Liu, F.~C. Li, C.~F. Lan,
\newblock \emph{Sci. Rep.} \textbf{2017}, \emph{7}, 1 12068.

\bibitem{Zhao2018a}
X.~Zhao, J.~Schalch, J.~Zhang, H.~R. Seren, G.~Duan, R.~D. Averitt, X.~Zhang,
\newblock \emph{Optica} \textbf{2018}, \emph{5}, 3 303.

\bibitem{Choi2019}
W.~J. Choi, G.~Cheng, Z.~Huang, S.~Zhang, T.~B. Norris, N.~A. Kotov,
\newblock \emph{Nat. Mater.} \textbf{2019}, \emph{18}, 8 820.

\bibitem{Liu2018a}
H.~Liu, J.~Lu, X.~R. Wang,
\newblock \emph{Nanotechnology} \textbf{2018}, \emph{29}, 2 024002.

\bibitem{Okimura2005}
K.~Okimura, N.~Kubo,
\newblock \emph{Jpn. J. Appl. Phys.} \textbf{2005}, \emph{44} L1150.

\bibitem{Wang2015b}
D.~Wang, L.~Zhang, Y.~Gu, M.~Q. Mehmood, Y.~Gong, A.~Srivastava, L.~Jian,
  T.~Venkatesan, C.-W. Qiu, M.~Hong,
\newblock \emph{Sci. Rep.} \textbf{2015}, \emph{5}, 1 15020.

\bibitem{Wang2016}
D.~Wang, L.~Zhang, Y.~Gong, L.~Jian, T.~Venkatesan, C.-W. Qiu, M.~Hong,
\newblock \emph{IEEE Photon. J.} \textbf{2016}, \emph{8}, 1 5500308.

\bibitem{Nouman2018a}
M.~T. Nouman, J.~H. Hwang, M.~Faiyaz, K.-J. Lee, D.-Y. Noh, J.-H. Jang,
\newblock \emph{Opt. Express} \textbf{2018}, \emph{26}, 10 12922.

\bibitem{Nakata2016}
Y.~Nakata, Y.~Urade, K.~Okimura, T.~Nakanishi, F.~Miyamaru, M.~W. Takeda,
  M.~Kitano,
\newblock \emph{Phys. Rev. Appl.} \textbf{2016}, \emph{6}, 4 044022.

\bibitem{Nakata2019c}
Y.~Nakata, K.~Fukawa, T.~Nakanishi, Y.~Urade, K.~Okimura, F.~Miyamaru,
\newblock \emph{Phys. Rev. Appl.} \textbf{2019}, \emph{11}, 4 044008.

\bibitem{Nakanishi2020}
T.~Nakanishi, Y.~Nakata, Y.~Urade, K.~Okimura,
\newblock \emph{Appl. Phys. Lett.} \textbf{2020}, \emph{117}, 9 091102.

\bibitem{Ding2018}
F.~Ding, S.~Zhong, S.~I. Bozhevolnyi,
\newblock \emph{Adv. Opt. Mater.} \textbf{2018}, \emph{6}, 9 1701204.

\bibitem{Song2018}
Z.~Song, K.~Wang, J.~Li, Q.~H. Liu,
\newblock \emph{Opt. Express} \textbf{2018}, \emph{26}, 6 7148.

\bibitem{Li2019}
X.~Li, S.~Tang, F.~Ding, S.~Zhong, Y.~Yang, T.~Jiang, J.~Zhou,
\newblock \emph{Sci. Rep.} \textbf{2019}, \emph{9}, 1 5454.

\bibitem{Wang2019}
S.~Wang, C.~Cai, M.~You, F.~Liu, M.~Wu, S.~Li, H.~Bao, L.~Kang, D.~H. Werner,
\newblock \emph{Opt. Express} \textbf{2019}, \emph{27}, 14 19436.

\bibitem{Song2020a}
Z.~Song, A.~Chen, J.~Zhang,
\newblock \emph{Opt. Express} \textbf{2020}, \emph{28}, 2 2037.

\bibitem{He2020}
H.~He, X.~Shang, L.~Xu, J.~Zhao, W.~Cai, J.~Wang, C.~Zhao, L.~Wang,
\newblock \emph{Opt. Express} \textbf{2020}, \emph{28}, 4 4563.

\bibitem{Song2020}
Z.~Song, J.~Zhang,
\newblock \emph{Opt. Express} \textbf{2020}, \emph{28}, 8 12487.

\bibitem{Zhang2020}
Y.~Zhang, P.~Wu, Z.~Zhou, X.~Chen, Z.~Yi, J.~Zhu, T.~Zhang, H.~Jile,
\newblock \emph{IEEE Access} \textbf{2020}, \emph{8} 85154.

\bibitem{Li2020}
H.~Li, J.~Yu,
\newblock \emph{Opt. Express} \textbf{2020}, \emph{28}, 17 25225.

\bibitem{Zhang2021a}
H.~Zhang, F.~Ling, B.~Zhang,
\newblock \emph{Opt. Mater.} \textbf{2021}, \emph{112} 110803.

\bibitem{Ren2021}
Y.~Ren, T.~Zhou, C.~Jiang, B.~Tang,
\newblock \emph{Opt. Express} \textbf{2021}, \emph{29}, 5 7666.

\bibitem{Liang2021}
J.~Liang, K.~Zhang, D.~Lei, L.~Yu, S.~Wang,
\newblock \emph{Appl. Opt.} \textbf{2021}, \emph{60}, 11 3062.

\bibitem{Liu2021a}
Y.~Liu, R.~Huang, Z.~Ouyang,
\newblock \emph{Opt. Express} \textbf{2021}, \emph{29}, 13 20839.

\bibitem{Shabanpour2020}
J.~Shabanpour, S.~Beyraghi, A.~Cheldavi,
\newblock \emph{Sci. Rep.} \textbf{2020}, \emph{10}, 1 8950.

\bibitem{Ren2021a}
B.~Ren, Y.~Feng, S.~Tang, L.~Wang, H.~Jiang, Y.~Jiang,
\newblock \emph{Opt. Express} \textbf{2021}, \emph{29}, 11 17258.

\bibitem{Luo2020}
J.~Luo, X.~Shi, X.~Luo, F.~Hu, G.~Li,
\newblock \emph{Opt. Express} \textbf{2020}, \emph{28}, 21 30861.

\bibitem{Jia2018}
Z.-Y. Jia, F.-Z. Shu, Y.-J. Gao, F.~Cheng, R.-W. Peng, R.-H. Fan, Y.~Liu,
  M.~Wang,
\newblock \emph{Phys. Rev. Appl.} \textbf{2018}, \emph{9}, 3 034009.

\bibitem{Liu2019}
M.~Liu, Q.~Xu, X.~Chen, E.~Plum, H.~Li, X.~Zhang, C.~Zhang, C.~Zou, J.~Han,
  W.~Zhang,
\newblock \emph{Sci. Rep.} \textbf{2019}, \emph{9}, 1 4097.

\bibitem{Shimatsu2010}
T.~Shimatsu, M.~Uomoto,
\newblock \emph{ECS Trans.} \textbf{2010}, \emph{33}, 4 61.

\bibitem{Liu2018}
X.~Liu, X.~Chen, E.~P.~J. Parrott, C.~Han, G.~Humbert, A.~Crunteanu,
  E.~Pickwell-MacPherson,
\newblock \emph{APL Photonics} \textbf{2018}, \emph{3}, 5 051604.

\bibitem{Nag2008}
J.~Nag, R.~F. {Haglund Jr},
\newblock \emph{J. Phys. Condens. Matter} \textbf{2008}, \emph{20}, 26 264016.

\bibitem{Ko2008}
C.~Ko, S.~Ramanathan,
\newblock \emph{J. Appl. Phys.} \textbf{2008}, \emph{104}, 8 086105.

\bibitem{Grischkowsky1990}
D.~Grischkowsky, S.~Keiding, M.~van Exter, C.~Fattinger,
\newblock \emph{J. Opt. Soc. Am. B} \textbf{1990}, \emph{7}, 10 2006.

\bibitem{Laman2008}
N.~Laman, D.~Grischkowsky,
\newblock \emph{Appl. Phys. Lett.} \textbf{2008}, \emph{93}, 5 051105.

\bibitem{Pozar2011}
D.~M. Pozar,
\newblock \emph{{Microwave Engineering}},
\newblock John Wiley \& Sons, Hoboken, New Jersey, 4th edition, \textbf{2011}.

\bibitem{Xu2016a}
H.-X. Xu, S.~Sun, S.~Tang, S.~Ma, Q.~He, G.-M. Wang, T.~Cai, H.-P. Li, L.~Zhou,
\newblock \emph{Sci. Rep.} \textbf{2016}, \emph{6} 27503.

\bibitem{Zhang2021}
D.-Q. Zhang, F.-Z. Shu, Z.-W. Jiao, H.-W. Wu,
\newblock \emph{Opt. Express} \textbf{2021}, \emph{29}, 5 7494.

\bibitem{Mian2015}
M.~S. Mian, K.~Okimura, J.~Sakai,
\newblock \emph{J. Appl. Phys.} \textbf{2015}, \emph{117}, 21 215305.

\bibitem{Kim2019}
Y.~Kim, P.~C. Wu, R.~Sokhoyan, K.~Mauser, R.~Glaudell, G.~{Kafaie Shirmanesh},
  H.~A. Atwater,
\newblock \emph{Nano Lett.} \textbf{2019}, \emph{19}, 6 3961.

\end{thebibliography}
\end{document}